\documentstyle[preprint,epsf,aps]{revtex}
\begin{document}
\title{  Quantum entanglement swapping with spontaneous 
parametric down conversion}
\author{ Xiang-Bin Wang\thanks{email: wang$@$qci.jst.go.jp}, 
B. S. Shi\thanks{email: shi$@$frl.cl.nec.co.jp}, 
A. Tomita\thanks{email: tomita$@$frl.cl.nec.co.jp}, and K. Matsumoto\thanks{email: keiji$@$qci.jst.go.jp}\\ 
Imai Quantum Computation and Information project, 
\\ERATO, Japan Sci. and Tech. Corp.,\\
Daini Hongo White Bldg. 201, 5-28-3, Hongo, Bunkyo, Tokyo 113-0033, Japan}
\maketitle 
\begin{abstract}
\begin{center} {\bf Abstract} \end{center}
Two remote parties that have never interacted
each other can be entangled through entanglement swapping operation
done by a third party. Currently existing entanglement swapping
experiments are done probabilistically by post-selection, i.e., 
once a successful swapping
is verified, the resultant entanglement is destructed.  
We propose a simple non-post-selection scheme to 
demonstrate the high quality quantum
entanglement swapping with the spontaneous parametric down conversion(SPDC) process.
 Our scheme only requires the
 normal photon detectors which only  distinguish the vacuum and non-vacuum
Fock states.
\end{abstract}
\begin{figure}
\begin{center}
\epsffile{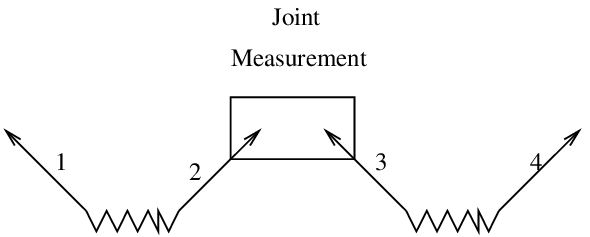}
\end{center}
\caption{
Schematic diagram of entanglement swapping. Initially, beam 1 and beam 2 are in EPR state, beam 3 and beam 4 are in another EPR state. After a joint measurement on beam 2 and 3 in Bell basis, beam 1 and beam 4 are projected into a Bell state.
}
\label{swap0}
\end{figure}

\noindent
{\it Introduction.}
Entanglement plays an important role in quantum mechanics. It is at the 
central role in the non-locality of quantum mechanics\cite{saku} including
Einstein-Podolsky-Rosen paradox, Bell's theorem and so on.
Entanglement is perhaps the most important resource in quantum computation
and information\cite{chuang}. To set up entanglement between particle A and
B, one may straightforwardly consider the method of collecting them
from the same source or of having them interact each other and then 
obtaining the entangled state of A and B after an appropriate non-trivial 
time evolution. 
However, one can also obtain the entanglement through  jointly 
measuring two particles (or light beams) in Bell basis as shown in Fig. \ref{swap0}.
 Note that this measurement
does not have to be done directly on particle A and B: When A is entangled
with particle A', B is entangled with B', a Bell measurement on A' and B'
will project A and and B into an entangled state.
That is to say,   
it is possible to entangle the two remote particles A and B
without any interaction by the method of 
{\it entanglement swapping}\cite{bennett,zu,knight}.
The first  entanglement swapping experiment was done by the Innsbruck group some years ago\cite{pant}. However, similar to the case of the quantum teleportation experiments\cite{bou,kimble0}, the result there is a post-selectin result:
once a successful entanglement swappingr is verified, the swapped entangled
state is destructed already (picture A in Fig. 2). 
Recently, the entanglement swapping has also been tested
in photon number space\cite{lom}. Again it is a post-selection test
unless  a  sophisticated photon detector to distinguish the one photon and two photons is used (picture B of Fig. 2).  Such a sophisticated photon detector is generally believed to be rather rare by our current technology therefore
it's not likely to really implement such a sophisticated
photon detector in the experiment.
 
\begin{figure}
\begin{center}
\epsffile{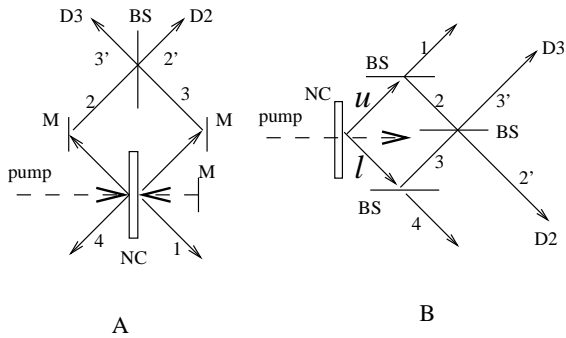}
\end{center}
\caption{ Currently existing entanglement swapping experiments 
are post-selection.
{\bf A.} The set-up in polarization space.
Because it's possible that beam 1 and 3 contains 2 pairs( or nothing) 
and beam 2 and 4
 contain nothing (or 2 pairs). 
To make sure beam 1 and 4 are entangled, one must also detect beam 1 and 4. 
{\bf B.} The set-up in vacuum-one photon space.
The photon detector here does not distinguish 1 photon or 2 photons.
When a detector is clicked, it's also possible that the beam contains 2 photons therefore the actual 
state for beam 1 and 4 is vacuum. To remove such types 
of events, one must also detect beam 1 and 4, therefore the state of beam 1 and 4 is destructed. NC: nonlinear crystal used in SPDC process. BS: beam splitter. M: mirror. D: photon detector.}
\label{inns}
\end{figure}

In this work, we report a very simple and robust non-post-selection
experimental scheme for the entanglement swapping based on the weakly entangled states initially.  Before going into details of our scheme, we examine the
post-selection nature of some existing experiments.
\\
{\it Post-selection nature of currently existing experiments.}
The existing experimental set-up in polarization space
is schematically shown in Fig.\ref{inns} A.
 An emitted pair will be
in the maximally entangled state in the polarization space if we only
collect the beam lights in crossing points of two emission cones\cite{kwiat}.
In the experiment, the pump light passes through the crystal twice.
The un-normalized total emitted state used there is
\begin{eqnarray}
|X\rangle_{1324}+|Y\rangle_{13} +|Y\rangle_{24}
\end{eqnarray}
where
\begin{eqnarray}
|X\rangle_{1324}=\left(|H\rangle_1 |V\rangle_3-|V\rangle_1 |H\rangle_3\right)
\left(|H\rangle_2 |V\rangle_4-|V\rangle_2 |H\rangle_4\right); 
\end{eqnarray} 
\begin{eqnarray}
|Y\rangle_{ij} = |2H\rangle_i|2V\rangle_j+ |2V\rangle_i|2H\rangle_j
-|HV\rangle_i|HV\rangle_j.
\end{eqnarray}
Only for the state $|X\rangle_{1234}$, the maximally entangled state
will be swapped to beam 1 and 4, provided that both D2 and D3 are clicked.
However, the initially emitted state contains the constitutes of $|Y\rangle$.
For this term, after D2 and D3 are both clicked, either beam 1 or beam
4 will contain nothing therefore they are not entangled. To overcome this,
the 4 fold detection is carried out in the experiment\cite{pant}. 
However,
this will destroy all swapped entanglement between 1 and 4.
The existing  experimental set-up in vacuum-one-photon space
 is shown in Fig.\ref{inns} B.
After passing through the beam splitter, the state is
\begin{eqnarray}
|00\rangle_{2'3'}|11\rangle_{14}+
\frac{1}{2}\left(|10\rangle_{2'3'}|\Psi^+\rangle_{14}+|01\rangle_{2'3'}|\Psi^{-}\rangle_{14}\right)+\frac{1}{\sqrt 2}|20\rangle_{2'3'}|00\rangle_{14}-
\frac{1}{\sqrt 2}|02\rangle_{2'3'}|00\rangle_{14}.
\end{eqnarray} 
Indeed, if beam $2'$ or beam $3'$ contains $exactly$ one photon, beam 1 and 4
will be maximally entangled. However, since the photon detectors do not distinguish one photon and two photon cases, the final result on beam 1 and beam 4
will be distorted by the constitute
$\frac{1}{\sqrt 2}|20\rangle_{2'3'}|00\rangle_{14}-
\frac{1}{\sqrt 2}|02\rangle_{2'3'}|00\rangle_{14}$. That is to say, whenever
one detector is clicked,  beam 1 and  4 is actually in 
a mixture  of a maximally entangled state and vacuum, instead of
a pure maximally entangled state. To overcome this,
beam 1 and beam 4 are also detected in the experiment\cite{lom}, again,
this post-selection operation will destroy all swapped entanglement between 1 and 4 whenever. 
\\
{\it Proposal for non-post-selection entanglement swapping.}
Consider the initial state
\begin{eqnarray}\label{ini}
|\Psi^-\rangle_{1234}=|\Psi^-\rangle_{12}\otimes |\Psi^-\rangle_{34}.
\end{eqnarray}
Obviously, none of particle 1,2 is
entangled with any particle of 3,4 at this stage.
However, if we jointly measure particle 2 and 3 in Bell basis, particle 1 and 4 will be projected to one of the 4 Bell state depending on the measurement 
result of particle 2 and 3. Explicitlyly, Eq.(\ref{ini}) can be recast into
the following form:
\begin{eqnarray}
|\Psi\rangle_{1234}=\frac{1}{2}\left(|\Psi^+\rangle_{23}|\Psi^+\rangle_{14}
-|\Psi^-\rangle_{23}|\Psi^-\rangle_{14} -|\Phi^+\rangle_{23}|\Phi^+\rangle_{14}
+|\Phi^-\rangle_{23}|\Phi^-\rangle_{14}\right)
\end{eqnarray}   
and $|\Phi^{\pm}\rangle_{i,j}
=\frac{1}{\sqrt 2}
\left(|0\rangle_i|0\rangle_j\pm |1\rangle_i|1\rangle_j\right)$, 
$|\Psi^\pm\rangle_{ij}
=\frac{1}{\sqrt 2}\left(|0\rangle_i|1\rangle_j
\pm |1\rangle_i|0\rangle_j\right)$.
This shows that whenever particle 2,3 is collapsed to a certain Bell state,
particle 1,4 is projected to the same Bell state therefore the maximal
entanglement 
between particle 1 and 4(Alice and Bob) is set up.
In Eq.(\ref{ini}) we have used  the initial state of product of
two antisymmetric states, actually, a product of arbitrary two maximally
entangled state will cause the similar result: after a joint measurement
to particle 2-3 in Bell basis, particle 1,4 will be projected to a maximally
entangled state. It has been shown in Ref.\cite{knight} that, even though
we start from a product  of non-maximally entangled states, we can still 
probabilistically obtain the maximal entanglement between 1 and 4 after the
joint measurement to particle 2,3. For example, we consider the following
initial state
\begin{eqnarray}
|\Psi'\rangle_{1234}
=|\theta\rangle_{12}|\theta\rangle_{34}
\end{eqnarray}  
and $|\theta\rangle_{ij}
=\cos \theta |0\rangle_i|0\rangle_j +\sin |1\rangle_i|1\rangle_j$.
This state cab be recast to
\begin{eqnarray} \nonumber
|\Psi'\rangle_{1234}=\frac{1}{ 2}\sin\theta\cos\theta \left(|\Psi^+\rangle_{23}
|\Psi^+\rangle_{14}-|\Psi^-\rangle_{23}|\Psi^-\rangle_{14}\right)+\\
\frac{1}{2}\left[|\Phi^+\rangle_{23}\left(\cos^2\theta |0\rangle_1|0\rangle_4
+\sin^2\theta |1\rangle_1|1\rangle_4\right) + 
|\Phi^-\rangle_{23}\left(\cos^2\theta |0\rangle_1|0\rangle_4
-\sin^2\theta |1\rangle_1|1\rangle_4\right)\right].
\end{eqnarray}
From this we can see that, even in the case that $\theta$ is very small, we
can still set up the maximal entanglement between particle 1 and 4
with a small probability through swapping operation. As we shall 
show it soon,
this small value of $\theta$ can be
an important advantage in a real experiment with imperfect entanglement
source and limitted power of practically existing devices. By making use of
the small value of $\theta$, one may test entanglement swapping without 
post-selection.
  
\begin{figure}
\begin{center}
\epsffile{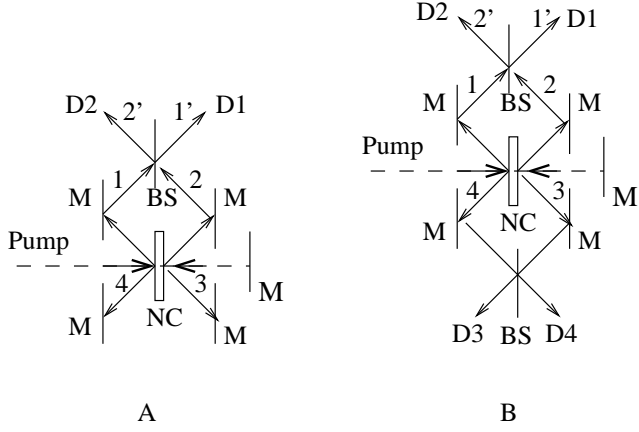}
\end{center}
\caption{{\bf A.} A schematic diagram for the experimental set-up of
 non-post-selection quantum entanglement swapping.
 Whenever we find the 
coincidence that  D$_1$  is fired(silent) and D$_2$ is  silent(fired), we have remotely prepared the entangled state of $|\Psi^+\rangle$($|\Psi^-\rangle$) on beam 3 and 4.
{\bf B.} Phase information verification of the entanglement swapping. The fact that detectors D$_1$(D$_2$) and D$_3$(D$_4$)  will be always both fired(silent) or both silent(fired) verifies the maximal entanglement of the state prepared by the entanglement swapping. }
\label{swap}
\end{figure}

Our strategy is to build up the entanglement between Alice and Bob, given two
copies of weakly entangled state, e.g.
\begin{eqnarray}
|\chi\rangle=\frac{1}{\sqrt{1+\epsilon^2}}(|00\rangle+\epsilon |11\rangle)
\end{eqnarray}  
which weakly entangles  Alice and Clare, Bob and Clare respectively.
After a Bell type measurement in Clare$'$s subspace and certain specific result( the coincidence event) is observed, we believe we have create a state $\rho_{AB}$  between Alice and Bob satisfying one of the following two equations
\begin{eqnarray}
\langle\Psi^{\pm}|\rho_{AB}|\Psi^{\pm}\rangle \sim 1
\end{eqnarray}
and $|\Psi^\pm\rangle=\frac{1}{\sqrt 2}(|10\rangle \pm |01\rangle)$.
Our first experimental scheme is schematically shown in fig.\ref{swap} A.
The spontaneous parametric down conversion process may happen after the pump light
passes through the nonlinear crystal. The total state for all the four beams can be written 
in the following form(in a good approximation)
\begin{eqnarray}
|\chi_0\rangle 
= \frac{1}{1+|\tau|^2}( |0\rangle_1 |0\rangle_4+\tau |1\rangle_1 |1\rangle_4 )
(|0\rangle_2 |0\rangle_3+\tau |1\rangle_2 |1\rangle_3)
\end{eqnarray}
where the subscripts indicate the specific modes(subspaces) and $|\tau|<<1$. Here beam 4 and beam 3
are  interpreted as in the subspaces of Alice and Bob respectively, while beam 1 and
beam 2 both belong to the subspace of Clare. In the experiment we should arrange the optical paths of beam 2 and beam 3 appropriately so that they reach the beam splitter simultaneously. $Either$ of the following two events 
 indicates a successful creation of a maximal entanglement in beam 4 and 3.

{\bf Event 1}. Detector D$_1$ is fired and D$_2$ is silent. Such an event
indicates that an entangled state $|\Psi^+\rangle$ is prepared on beam 3 and beam 4.

{\bf Event 2}. Detector D$_2$ is fired and D$_1$ is silent. Such an event
indicates that an entangled state $|\Psi^-\rangle$ is prepared on beam 3 and beam 4.

We denote $U_B$ as the time evolution operator of
our beam splitter. We assume the following
  properties for the (balanced) beam splitters in the
Schrodinger picture\cite{campos0,wang}
   \begin{eqnarray}
U_B(a_1^\dagger, a_2^\dagger)U_B^\dagger=\frac{1}{\sqrt 2}(a_1^\dagger, a_2^\dagger)\left(
\begin{array}{cc}1 & 1 \\ 1 & -1\end{array}
\right).\label{db}
\end{eqnarray}
Here $a_1^\dagger$ and $a_2^\dagger$ are creation operators of mode 1 and
mode 2 respectively. There are many other forms of balanced beam splitters,
but all of them will essentially cause the same result on entanglement
swapping by  our scheme therefore in this paper we only consider the one
defined above. 
Note that here we are using the Schrodinger picture and we 
simply distinguish different mode around the beam splitter by the propagation direction only\cite{wang}. For example, beam 1 and beam 1' are in the same mode
but different state due to the nontrivial time evolution in the two mode space 
caused by the beam splitter.
Using all this, we know that the total state after the beam solitter is 
\begin{eqnarray}
|\chi_1\rangle=U_B|\chi_0\rangle
=\frac{1}{1+|\tau|^2}\left[ |0000\rangle+\frac{\tau}{\sqrt 2} |10\rangle(|10\rangle+|01\rangle)-\frac{\tau}{\sqrt 2}|01\rangle(|10\rangle - |01\rangle)+\tau^2 U_B|1111\rangle
\right].\label{im}
\end{eqnarray} 
In the above equation we have omitted all  subscripts for the mode indicators. In
all the state vectors in the format of $|wxyz\rangle$ or in the format of 
$|wx\rangle|yz\rangle$, we always assume that
 the symbol  in the first, second, third and fourth position
from the left to the right are for the quantum state in beam 1', 2', 3 4 respectively.
Equation(\ref{im}) shows that once event 1 or event 2 happens, we have obtained the
state $|\Psi^-\rangle=\frac{1}{\sqrt 2}(|1\rangle_3|0\rangle_4 - |0\rangle_3
|1\rangle_4)$ or $|\Psi^+\rangle=\frac{1}{\sqrt 2}(|1\rangle_3|0\rangle_4 + |0\rangle_3
|1\rangle_4)$
with a probability of $p=1-|\tau |^2 $.   Suppose $|\tau|^2=10^{-3}$,
this probability is  about $99.9\%$
Note that the photon detectors used here  need not be capable of  distinguishing
 one photon and two photons. Moreover, we even don$'$t have to worry about the efficiency of the photon detectors due to the very small value of $|\tau|^2$. 

\begin{figure}
\begin{center}
\epsffile{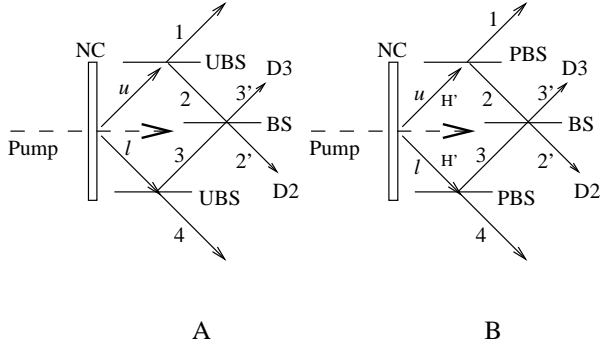}
\end{center}
\caption{ {\bf A.} An alternative scheme for non-post-selection entanglement swapping.
UBS: unbalanced beam splitter; BS: balanced beam splitter. 
After the clicking of either D2 or D3, beam 1
and beam 4 are in the maximally single photon entangled state in the two level
space of vacuum-one-photon state. {\bf B.} The unbalanced beam splitters
in A can be replaced by polarizing beam splitters.} 
\label{sgj1}
\end{figure}

In an experimental test, we need to verify that beam 3 and beam 4 are indeed entangled after we observed the event 1 or event 2 successfully. Doing so is quite simple. First we check the probability distribution. The state $|\Psi^\pm\rangle$
will give the equal classical probability of one photon on beam 4 and beam 3.
To check this, we only need to place extra  photon detectors on  those two beams.
We then check the phase information which distinguishes $|\Psi^+\rangle$ and
$|\Psi^-$.
To do so we just let beam 3 and beam 4 pass through   another 
balanced beam splitter defined by eq.(\ref{db}). For state$|\Psi^+\rangle$, we always find a photon left to the beam splitter, while for the state $|\Psi^-\rangle$,  we always find a photon right to the beam splitter. That is to say, in order to verify the phase information, we just observe the coincidence that D$_1$ is always fired together with D$_3$, and that D$_2$ is always
fired together with D$_4$ in fig.\ref{sgj1} B. Due to the limitted efficiency of photon detectors, normally we cannot always observe the theoretically
expected coincidence. That is to say, whenever D1 is clicked, D3 clicks only
in a probability of $\eta$, where $\eta$ is the photon detector efficiency. 
But the fact that whenever D1 clicks, D4 never( or rarely) clicks will be a strong evidence of phase coherence between beam 3 and beam 4. 

In the above scheme, the pump light there has to pass through the nonlinear crystal twice.
This may increase technical difficulty in synchronization. To avoid this,
we also propose the following alternative scheme shown in Fig. \ref{sgj1}
where the pump light only pass through the nonlinear crystal once. 
 The 
 unbalanced beam splitter is almost transparent.
Its time evolution operator satisfies
\begin{eqnarray}
U(a_1^\dagger, a_2^\dagger)U^\dagger=\frac{1}{\sqrt {1+\epsilon^2}}
(a_1^\dagger, a_2^\dagger)\left(
\begin{array}{cc}1 &  \epsilon \\ \epsilon & -1\end{array}
\right).\label{udb}
\end{eqnarray}  
This shows that, in the case that one pair $|1\rangle_u|1\rangle_l$
is emitted from the nonlinear crystal, 
 after passing through unbalanced beam splitters, the initial state is evolved to 
\begin{eqnarray}\label{inin}
|\chi'\rangle_{1234}=\frac{1}{1+\epsilon^2}\left(
|1\rangle_1|0\rangle_2 + \epsilon |0\rangle_1|1\rangle_2
\right)\otimes \left(
\epsilon |1\rangle_3|0\rangle_4 + |0\rangle_3|1\rangle_4
\right).
\end{eqnarray}
This can be recast to
\begin{eqnarray}\label{rec}
|\chi'\rangle_{1234}=\frac{1}{1+\epsilon^2}\left[
|1\rangle|0\rangle |0\rangle|1\rangle + 
\epsilon \left( |0\rangle|1\rangle|0\rangle|1\rangle
+ |1\rangle|0\rangle|1\rangle|0\rangle\right)
+ \epsilon^2|0\rangle|1\rangle
 |1\rangle|0\rangle
\right].
\end{eqnarray}
where we have omitted the subscripts for all beams. We just keep in the mind
that for each term the subscripts are from 1 to 4 from left to right.
One can easily show that whenever  D2 or D3 is clicked, beam 1,4 must be
projected onto a Bell state.
The first term in the right hand side of Eq.(\ref{rec}) will never cause any
clicking because beam 2 and beam 3 contain nothing there. The last term
can cause the clicking of D2 or D3, but the probability is very small because
the value of $\epsilon^2$ is much smaller than $\epsilon$. Therefore one 
need only consider the middle term, i.e., the term with a 
factor of $\epsilon$.
The second term can be written in the equivalent form of
\begin{eqnarray}
\frac{\epsilon}{2}\left(|\Psi^+\rangle_{23}|\Psi^+\rangle_{14}
-|\Psi^-\rangle_{23}|\Psi^-\rangle_{14}\right).
\end{eqnarray} 
We know that state $|\Psi^+\rangle_{23}$ will cause detector D2 being clicked and
state $|\Psi^-\rangle_{23}$ will cause the detector D3 being clicked.
Therefore, in the setup given by Fig.\ref{sgj1}, whenever D2 is clicked,
beam 1,4 have been projected to $|\Psi^+\rangle_{14}$;
 whenever D3 is clicked,
beam 1,4 have been projected to $|\Psi^-\rangle_{14}$.
We can choose to use the polarizing beam splitters (PBS) instead of the 
unbalanced beam splitters there.  We need only to rotate
beam u and beam l appropriately therefore the polarization in both
beams are a bit deviate from the horizontal one, i.e. initially we produce 
a state of 
\begin{eqnarray}
|H'\rangle_u\otimes |H'\rangle_l = \frac{1}{1+\epsilon^2}(|H\rangle_u+\epsilon |V\rangle_u)\otimes (|H\rangle_l+\epsilon |V\rangle_l).
\end{eqnarray}
Now the two unbalanced beam splitters are replaced by two polarizing beam
 splitters. The state for beam 1,2,3 and 4 will be identical to that given by Eq.(\ref{inin}) therefore all the rest results are the same.\\
{\it Concluding remark.}
In summary, we have given a $simple$ proposal to do the 
quantum entanglement swapping experiment without postselection. 
Since the post-selection experiments with similar or more complicated
technical setup have been carried out already\cite{pant,lom},
we  believe our scheme can be carried out easily with the current technology. We don't know how to scale the method to the case of many entangled pairs.\\
{\bf Acknowledgement:} 
We thank Prof Imai H for support.

\end{document}